\documentstyle[12pt,./aaspp4]{article}

\newcommand{\pyr}{\mbox{${\rm yr}^{-1}$}}

\newcommand{\msun}{\mbox{${\rm M}_\odot$}}
\newcommand{\msunyr}{ \mbox{ ${\rm M}_\odot \,{\rm yr}^{-1}$ } }
\newcommand{\lsun}{\mbox{${\rm L}_\odot$}}
\newcommand{\rsun}{\mbox{${\rm R}_\odot$}}
\newcommand{\kms}{\mbox{${\rm km~s}^{-1}$}}

\def\apgt{\ {\raise-.5ex\hbox{$\buildrel>\over\sim$}}\ }
\def\aplt{\ {\raise-.5ex\hbox{$\buildrel<\over\sim$}}\ }

\def\apgt{\ {\raise-.5ex\hbox{$\buildrel>\over\sim$}}\ }
\def\aplt{\ {\raise-.5ex\hbox{$\buildrel<\over\sim$}}\ }
\lefthead{Simon F.\ Portegies Zwart}
\righthead{OB runaways from supernova ejections} 

\begin{document}

\title{The characteristics of high velocity O and B stars which are
       ejected from supernovae in binary systems}

\author{Simon F.\ Portegies Zwart\footnote{Hubble Fellow}}
\affil{Department of Astronomy,
		 Boston University,
		 725 Commonwealth Ave.,
		 Boston, MA 02215, USA}
\authoremail{spz@komodo.bu.edu}


\date{Received; accepted: }

\begin{abstract}
We perform binary population synthesis calculations to study the
origin and the characteristics of runaway O and B stars which are
ejected by the supernova explosion of the companion star in a binary
system.  The number of OB runaways can be explained from supernova
ejections only if: high mass stars are preferentially formed in
binaries, the initial mass ratio distribution is strongly peaked to
unity and stars are rejuvenated to zero age upon accretion of mass
from a companion star.  Taking these requirements into consideration
we conclude that at most 30\% of the runaway O stars but possibly all
runaway B stars obtain high velocities due to supernovae in evolving
binaries.  Stars which obtain high velocities via supernova ejections
have the following characteristics: 1) at least 10\% of the high
velocity B stars and half the O stars have a mass greater than the
turn off mass of the cluster in which they are born and would be
observed as blue stragglers in the parent cluster, 2) their equatorial
rotational velocities are proportional to their space velocity and 3)
between 20\% and 40\% of the runaways have neutron star companions but
less than 1\% are visible as radio pulsars in part of the orbit.
\end{abstract}


\section{Introduction}\label{sect:introduction}

OB runaways are among the most massive stars with high velocities
($\apgt 25$\,\kms) or are located far ($\apgt 100$\,pc) from their
estimated birth place compared to other O and B stars in the galactic
disc (Gies \& Bolton 1986; Gies
1987).\nocite{1986ApJS...61..419G}\nocite{1987ApJS...64..545G} (The
terms OB runaway and runaway O and B stars will be used
interchangeably.)

The first discovered OB runaways, the B0\,V star $\mu$\,Col and the
O9.5\,V star AE Au (Blaauw \& Morgan 1954)\nocite{1954ApJ...119..625B}
started a half century debate about their characteristics and the
origin of the, sometimes inferred, high velocities.
  
The definition of a runaway varies between researchers. Blaauw (1961)
argues that only stars with a velocity in excess of 40\kms\, should be
called runaways, Stone (1991) on the other hand, defines a runaway as
a star which belongs to a separate group of stars which follows a
Maxwellian velocity distribution with a higher mean.  Although Stones'
(1991)\nocite{1991AJ....102..333S} definition is theoretically most
distinctive, the definition of a velocity cut-off is the most
practical.  We define a runaway as a star with velocity $v>25$\,\kms,
and the specific frequency of runaways $f_{\rm run}$ as the number of
runaways with mass $M$ as fraction of all stars with the same mass.

The specific frequency of runaways among O and B stars decreases
steeply from $f_{\rm run} \simeq 20$\% among the earliest O stars to
about 2.5\% among early B stars and even smaller frequencies among
later B1 to B5 stars (Blaauw 1961). Stetson
(1981)\nocite{1981AJ.....86.1882S} found that $\apgt 30$\% of A stars
and almost all F stars have high velocities.  These velocities may be
the result of the heating of the Galactic disc by molecular clouds and
Stone (1991) concludes that the fraction of true runaways among A
stars is $\sim 0.3$\% (less than 1\%).

Many runaways are characterized by high rotational velocities (Conti
\& Ebbets 1977)\nocite{1977ApJ...213..438C}. Some runaways can be
traced back to the association in which they were born.  The runaway
stars $\zeta$\, Oph, AE Aur, $\mu$\, Col and 53 Ari were blue
stragglers if placed back in the association from which they are
ejected (Blaauw 1993; Hoogerwerf et al.\,
2000).\nocite{1993msli.conf..207B}\nocite{2000Hoogerwerf_etal}
Accurate distances and proper motions of most of the runaway O and B
stars may have to wait until the FAME (Full Sky Astrometric Mapping
Explorer) mission. This satellite will measure accurate positions and
velocities of O and B star to a distance of several kiloparsec.

The observed frequency of single line spectroscopic binaries among OB
runaways $\aplt 19$\% (Gies 1987) where among other O stars this is
between 40\% (Garmany \& Conti 1980)\nocite{1980IAUS...88..163G} and
60\% (Conti et al.\, 1977).\nocite{1977ApJ...214..759C} Sayer et al.\,
(1996, see also Philp et al.\, 1996) conclude that between 20\% and
40\% of the OB runaways may have a pulsar companion but most are
unobservable.

A number of scenarios have been proposed to explain the origin of OB
runaway stars. The following two ejection mechanisms are most
promising:
\begin{itemize}
\item[$\bullet$] A star is ejected from a close binary at the moment its
	         companion explodes in a supernova (Blaauw
	         1961).\nocite{blaauw_1961} 
\item[$\bullet$] A star is ejected in a dynamical 3- or 4-body
	         encounter (Poveda et al.\, 1967).\nocite{Poveda_1967}
\end{itemize}

In this paper we study the first mechanism using binary population
synthesis.  Previous population synthesis studies claim that all OB
runaways are readily explained by the supernova ejection mechanism.
Stone (1982, see also Leonard \& Dewey 1993; De Donder et al.\, 1997)
conclude that about half the O stars have a high space velocity and
that 20\% to 40\% of these may have a compact binary companion in an
eccentric
orbit.\nocite{1993lhls.work..239L}\nocite{1982AJ.....87...90S} High
rotational velocities and high Helium abundances are generally
explained by assuming that the pre-supernova binary has experienced a
phase of mass transfer. The star may be rejuvenated during such a
phase which may explain why some observed runaways are blue stragglers
(Blaauw 1993).  The observation of bow shocks around about a dozen
Wolf-Rayet stars suggests that their velocities are also high.  The
recently observed bow shocks around the X-ray binary Vela X-1 (Kaper
et al.\, 1997)\nocite{1997ApJ...475L..37K}
and EZ Cma (van Buren et al.\, 1995)\nocite{1995AJ....110.2914V}
indicates that these binaries have high space motions, which may be
obtained in the supernova in which the compact object formed (van
Rensbergen et al.\, 1996).\nocite{1996A&A...305..825V}

Opposing these are the claims that OB runaways can be readily
explained by dynamical ejection from young and dense star clusters
(Aarseth 1974; Gies \& Bolton 1986; Leonard and Duncan
1988).\nocite{1974A&A....35..237A}\nocite{1988AJ.....96..222L} The
high velocities of some observed runaways can be explained in this
process (Conlon et al.\, 1990; Leonard
1991).\nocite{1991AJ....101..562L}\nocite{1990A&A...236..357C} And
high Helium abundances, rotational velocities and the fact that some
are blue stragglers could be explained from star which are ejected in
a binary-binary encounter in which the ejected star first merges with
one of the other stars (Leonard 1995).\nocite{1995MNRAS.277.1080L} The
runaway pair $\mu$\, Col and AE Aur is most easily explained in this
way as $\sim 2.7$\,Myaer ago both stars may have been ejected in
opposite direction from the association Ori OB1 (Blaauw \& Morgan
1954; Hoogerwerf et al.\, 2000), indicating a dynamical origin rather
than a supernova ejection. Other evidence for dynamical ejection is
provided by the existence of two double-lined spectroscopic binaries
among the sample of known runaways (Gies \& Bolton 1986).

Supporters of either mechanism claim to be able to explain the
observable characteristics of O and B runaways and their specific
frequencies. In this papers I discuss the characteristics of massive
stars which obtained a high velocity from a supernova in a close
binary system. This paper is organized as follows. Section\,2
describes the binary evolution program used for the calculations, the
results of the calculations are described in section\,3 and we discuss
these in section\,4. Section\,5 sums up.

\section{Methods}
We use the binary population synthesis code {\sf SeBa}\footnote{The
name {\sf SeBa} is adopted from the Egyptian word for `to teach', `the
door to knowledge' or `(multiple) star'. The exact meaning depends on
the hieroglyphic spelling.} (Portegies Zwart \& Verbunt
1996)\nocite{pzv96} with adjustments as in model B from Portegies
Zwart \& Yungelson (1998).\nocite{pzy98}

The binaries are initialized as follows: The mass of the most massive
star (primary) is selected from the initial mass function for the
Solar neighborhood as described by Scalo (1986;
1998).\nocite{scalo86}\nocite{1998simf.conf..201S} The masses of the
companions are randomly selected between 0.1\,\msun\, and the primary
mass. Then the other binary parameters are determined.  Binary
eccentricities are selected from the thermal distribution between 0
and 1.  Orbital separations $a$ are selected with equal probability in
$\log a$ between 1\,\rsun (or Roche-lobe contact whichever is larger)
and an upper limit of $10^6$\,\rsun\, (about 0.02\,pc).
Table\,\ref{Tab:Binit} gives an overview of the various distribution
functions from which stars and binaries are initialized.  In
sect.\,\ref{sect:discussion} we study the effect of changing these
assumptions.

\placetable{Tab:Binit}

We will only mention some of the model features of {\sf SeBa} (for
details see Portegies Zwart \& Verbunt 1996).  Single stars between
$\sim 9$ and 25\,\msun\, explode in a supernova to form a neutron
star. Less massive and more massive stars become white dwarfs and
black holes, respectively. Neutron stars receive a velocity kick upon
birth.  Following Hartman et al.\, (1997),\nocite{1997A&A...322..127H}
we assume the distribution for isotropic kick velocities
\begin{equation}
P(u)du = {4\over \pi} \cdot {du\over(1+u^2)^2},
\label{eqkick}\end{equation}
with $u=v/\sigma$ and $\sigma = 600 \,\,\mbox{km}\,\mbox{s}^{-1}$.

A star which receives mass from its companion in a close binary system
may be rejuvenated in the process. The accreting star is rejuvenated
with the product of its current age and the fraction of mass gained in
the mass transfer process (see Portegies Zwart \& Verbunt
1996 for details). Rejuvenation is thus less effective early on in the
evolution of the star (near the zero-age main sequence) but becomes
more effective when the accretor approaches the terminal-age main
sequence.

\section{Results}

We initialize $10^6$ binaries with a primary mass between $M=1$\,\msun\,
and 100\,\msun. Mass transfer and stellar winds may affect the masses
of both stars and the binary parameters. The binaries are evolved
until the first supernova, which is the only way to obtain a velocity
$v>0$. 

Figure\,\ref{fig:Mrunfrac} gives, as a function of mass, the number of
stars which are ejected in a supernova as a fraction of the
initialized primaries with the same mass.  The fraction of stars in
which the companion experiences a supernova is a steep function of
mass (bullets in Figure\,\ref{fig:Mrunfrac}).  Binaries with a primary
star $M \aplt 9$\,\msun\, do not produce runaways because the primary
is not massive enough to experience a supernova. Runaways with a mass
$m \aplt 9$\,\msun\, can still be produced but only from binaries in
which the primary was massive enough to experience supernova.  The
filled triangles in figure\,\ref{fig:Mrunfrac} shows the specific
frequency of runaways (according to our definition: stars which
received a velocity in excess of 25\,\kms). The fraction of runaways
is much smaller than the fraction of stars in which the companion
experiences a supernova; many supernova explosions occur in rather
wide binaries resulting in small ejection velocities.

Runaway stars may obtain high velocities after a complicated evolution
as member of a close binary.  This initial 'pre runaway' phase lasts
until the primary star with mass $M$ explodes in a supernova: $t_{\rm
sn}(M)$.  The runaway has thus spends $t_{\rm sn}(M)$ as a
non-runaway. Before the supernova, however, the binary may have
experienced a phase of mass transfer, rejuvenating the accretor with
$dt_{\rm Bss}$ i.e., the time gained on the main sequence due to
rejuvenation.  The lifetime of the star then becomes $t_{\rm
ms}(m) + dt_{\rm Bss}$ and the total time it spends as a runaway is
\begin{equation}
	t_{\rm run} = t_{\rm ms}(m)-t_{\rm sn}(M) + dt_{\rm Bss}.
\end{equation}\label{Eq:trun}
All other stars, including the primaries of the binaries which do not
gain high velocities because they do not experience a supernova remain
visible for $t_{\rm ms}(M)$. (We neglect the post main-sequence
evolution for this population.)  The specific frequency of runaways
can then be defined by summing over all stars in the denominator and
all runaways in the nominator;
\begin{equation}
	f_{\rm run} \equiv { \sum_{\rm run} t_{\rm run} \over 
	         \sum_{\rm all} t_{\rm ms}(M) + \sum_{\rm run} dt_{\rm Bss}}.
\end{equation}\label{Eq:frun}
The extra time gained as a blue straggler appears in the
denominator as a correction factor.

The importance of the time lost as a non-runaway and the rejuvenation
of the accretor are demonstrated with the open triangles in
Fig.\,\ref{fig:Mrunfrac}.  The time gained by rejuvenation is not
enough to account for the time lost in the pre-supernova binary.  Only
$\sim 2$\% of the stars in the mass range from 6\,\msun\, to
30\,\msun\, are observable as runaways with velocities $>25$\,\kms.
These masses correspond at zero age to spectral type B3V and O6V for
the 6\,\msun\, and 30\,\msun\, star, respectively.

\placefigure{fig:Mrunfrac}

Figure\,\ref{fig:Mtrunaway} gives, as a function of mass, the fraction
of the lifetime that a star with mass $m$ spends as a runaway.
\begin{equation}
	\tau_{\rm run} \equiv {t_{\rm run}
		      	       \over 
			       t_{\rm ms}(M)  + dt_{\rm Bss}}.
\end{equation}\label{Eq:taurun}

The width of the distribution reflects the various accretion histories
of the stars. Rejuvenation is more effective if more mass is accreted
and if that happened later in the evolution of the accretor (near the
terminal-age main sequence). High mass stars spend a smaller fraction
of their lifetime as runaway.

\placefigure{fig:Mtrunaway}

Figure\,\ref{fig:MfBss} shows that $\apgt 95$\% of the runaways
(filled triangles) have accreted some mass and are therefore
rejuvenated. These stars appear younger than they really are, but his
only shows up if the turn-off mass of their parent association drops
below the mass of the runaway: if put back in the parent cluster the
runaway appears above the turn off as a blue straggler.  At any time
the fraction of blue straggler among runaway stars is only about 10\%
for B stars and about 40\% for O stars (open triangles). A higher mass
star spend a larger fraction of its runaway lifetime as a blue
straggler.

\placefigure{fig:MfBss}

Figure\,\ref{fig:Mv_all} shows the probability distribution (by
number) as a function of runaway velocity and mass.  High mass stars
tend to have smaller velocities than low mass stars; which is
consistent with observations (see Gies \& Bolton 1986).

\placefigure{fig:Mv_all}

Non-conservative mass transfer generally results in a shorter binary
orbit, since the lost mass carries angular momentum. The
runaways produced in such binaries tend to obtain higher speeds after
the supernova (see also van den Heuvel et al.\, 2000). Because the
runaway star has accreted less material it is evident that the highest
velocity runaways have lower anomal surface abundances, but higher
equatorial rotational velocities.  The latter is a result of the tidal
locking during a phase of mass transfer.


Figure\,\ref{fig:Mbinfrac} shows the fraction of binaries that
survives the first supernova (bullets) and the binary fraction among
runaways (triangles).  The binary fraction among runaways is about
20\% for late type B stars and increases to $\sim 40$\% for earlier
spectral types.

\placefigure{fig:Mbinfrac}

\placefigure{fig:Pv_bin}

Figure\,\ref{fig:Pv_bin} shows the probability distribution in orbital
period and space velocity of the binaries which survive the
supernova. Shorter period binaries often have higher space velocities
(see also van den Heuvel et al.\, 2000).

\section{Discussion}\label{sect:discussion}

We studied the characteristics of high velocity B and O stars using
binary population synthesis. The only way in which a star can obtain
a high velocity in the model is by the supernova explosion of the
companion star in a binary system. If the binary is dissociated in the
supernova a single neutron star (or black hole) and the rejuvenated
companion star are ejected, otherwise the runaway star will still be
accompanied by the compact object.  Independent whether the runaway
star is single or the member of a binary system it is likely to be a
blue straggler, rapidly rotating and the star may have a funny surface
abundance due to its history as an accreting star.

The observed frequency of runaways among O stars is $\sim 20$\%, much
higher than the $\sim 2$\% resulting from the model calculations.  The
$\sim 2$\% runaways among B stars from the model is on the low side
but not inconsistent with the observations reported by Blaauw (1993)
and Stone (1991).

These are serious discrepancies and may in part be solved by invoking
various observational selection effects and by adjusting initial
conditions and model parameters. We discuss each in turn.

Young O stars remain easily hidden in their parental clouds for a
substantial fraction of their lifetime.  The stars are born with low
velocities and gain speed at later age.  This means that for as long
as the star resides in the cloud it may be hard to observe. Hiding the
youngest --non runaway-- O stars enhances the fraction of star with
high velocity relative to low velocity objects.  Estimates based on
the lifetime of dense parental clouds indicates that the most massive
O stars spend less than about 1\, million years unobservable in the
optical. This is not enough to increase the fraction of runaways by
more than a few percent and can therefore not be responsible for the
order of magnitude discrepancy between the observed and computed
frequency of spectral type O runaways.

\subsection{Boosting the formation rates of type O runaways}
\label{sect:OBfrac}

The result of the population synthesis calculations are insensitive to
the initial mass function and the eccentricity distribution, but
selecting a different mass ratio distribution affects the results more
strongly.  Observations of high mass binaries favor a mass ratio
distribution which is somewhat peaked to unity (Garmany \& Conti 1980;
Hogeveen 1991).\nocite{1980IAUS...88..163G}\nocite{hog91}

An initial mass ratio distribution of the from $P(q) \propto q^3$ with
the same limits as adopted before results in an increase of the
fraction of runaway O stars and decreases the fraction of runaways
among later spectral types.  Figure\,\ref{fig:Mrunaway}\, (circles)
shows the results of these calculations with $10^6$\, binaries. The
fraction for O stars is still too low by at least a factor four, but
the observed trend is well explained.

Wide binaries do not contribute to the formation of runaways.
Decreasing the maximum orbital separation from $10^6$\,\rsun\, to
$10^4$\,\rsun\, increases the fraction of runaways by roughly 50\%
over the entire mass range.  The increase obtained by this change,
indicated by the vertical arrow in fig.\,\ref{fig:Mrunaway}, is
still not sufficient to explain the observed fraction of runaways
among O stars.

\placefigure{fig:Mrunaway}

Finally we may extend the lifetime of the runaway by assuming that
mass transfer causes the accretor to rejuvenate to zero age.
Combining this with the above discussed alternative mass-ratio
distribution results in a specific frequency of runaway stars
consistent with the observations (see $\bullet$ in
Fig.\,\ref{fig:Mrunaway}).  This indicates, however, that we still
under produce runaways by a factor two, as we assumed a binary
frequency of 100\% where the observations indicate that only $\sim
50$\% of the O stars reside in binaries (Garmany \& Conti 1980; Conti
et al.\,
1977).\nocite{1980IAUS...88..163G}\nocite{1977ApJ...214..759C}

We conclude from these arguments that at most 30\% of the runaway O
stars but most of the B stars can be explained from supernova
ejections.

\subsection{Hide and seek the pulsar}

The binary evolution model predicts a binary fraction among runaways
from 20\% for late B stars to $\sim 40$\% among early O stars.  These
binary frequencies are much higher than the observed $\aplt 8$\%
(Philp et al 1996; Sayer et al.\
1996).\nocite{1996AJ....111.1220P}\nocite{1996ApJ...461..357S} The
binary fraction among runaways is again presented in
Figure\,\ref{fig:MvisPbinfrac}, but we improved statistics by adding
the results of the calculation with the mass ratio distribution peaked
to unity from sect.\,\ref{sect:OBfrac} to our initial results.  (The
binary fraction among runaways is rather insensitive to the initial
mass ratio distribution.)

All runaway binaries contain a neutron star or a black hole.
There are a number of selection effects against finding a radio pulsar
as a companion to a massive star.  The low mass of a neutron star
compared to its accompanying O or B star makes it hard to observe the
periodicity in a radial velocity curve.  A radio pulsar dies after
about 10\,Myear. For runaways which remain visible for a shorter time
(for $M\apgt 15$\,\msun) this poses no limiting factor.  Low mass
stars, however, live much longer than the pulsar and the majority of
stars with $M\aplt8$\,\msun\, are therefore accompanied by an
unobservable --dead-- pulsar.

The triangles in Fig.\,\ref{fig:MvisPbinfrac} show the binary
fraction among runaways in which the neutron star is visible as a
pulsar.  We assumed here that the pulsar lives for 10\,Myear and
neglected the limited beaming fraction of the pulsar.  As expected,
low mass runaways are likely to be accompanied by a dead pulsar, where
the high mass stars do not suffer from this selection effect. The
chance of catching the neutron star as a pulsar among low mass
runaways is small, even though a large fraction of runaways is
accompanied by one.

The radio emission from a pulsar is easily absorbed by the mass lost
in the stellar wind of the O or B star.  The minimum rate of mass loss
in the stellar wind required to hide the pulsar can be estimated from
the optical depth of the stellar wind for free-free absorption (e.g.,
Eq.~16 in Illarionov \& Sunyaev 1975), we obtain
\begin{equation}
	{\dot M}_{\rm hide} \apgt 5.6\,10^{-13} (M+m)
		            \left[ a (1-e) 
			    \right]^{\frac{3}{2}} \,\,\; [\msunyr].
\label{eq:ppulsar}
\end{equation}
Here $M$, $m$, $a$ and $e$ are the primary and secondary masses (in
\msun), the semi major axis (in \rsun) and the orbital eccentricity of
the binary.  Here it is assumed that the wind of the
accompanying star is transparent at $\lambda = 75$\,cm and the
temperature of the stellar plasma is $10^4$\,K.  The estimate for
${\dot M}_{\rm hide}$ depends on the eccentricity; substitution of
$(1+e)$ for $(1-e)$ in Eq.~(\ref{eq:ppulsar}) gives the lower limit
for the mass loss rate given that the radio pulsar is only visible at
apocenter.

We estimate the mass loss rate for a main sequence star with mass $M$
(in \msun), luminosity $L$ (in \lsun) and radius $R$ (in \rsun) with
(Lamers 1981; Lamers et al.\,
1993)\nocite{1981ApJ...245..593L}\nocite{1993ApJ...412..771L}
\begin{equation}
	\log {\dot M}_{\rm wind} = 
	-4.83 - 4.58\log L - 0.87 \log R - 0.49 \log M.
\end{equation}
Here ${\dot M}_{\rm wind}$ is in \msun\,\pyr. Many of the runaways
have experienced a phase of mass transfer and may have enhanced mass
loss rates (Snow 1982),\nocite{1981ApJ...251..139S} which we do not
account for. If ${\dot M}_{\rm wind}$ exceeds ${\dot M}_{\rm hide}$
the pulsar is unobservable. The large circles in
Fig.\,\ref{fig:MvisPbinfrac} shows the number of binaries in which the
pulsar is visible through the stellar wind as fraction of all runaway
stars (single and binary). Some pulsar, however, are observable near
apocenter but hidden at pericenter (small circles in
Fig.\,\ref{fig:MvisPbinfrac}).

\placefigure{fig:MvisPbinfrac}

Between 30\% and 40\% of the high-mass (spectral type O) runaways are
accompanied by a radio pulsar, but 2\% to 4\% of these are visible
through the stellar wind of their companion. About 1\% of the early
type stars will therefore be observable as a radio source.  For early
spectral type B stars ($m\apgt 7$\,\msun) this fraction increases to
about 1.8\% mainly due to their lower mass loss rates, which makes it
easier for the pulsar to shine through. For later spectral type stars
the fraction drops again; mainly caused by the longer lifetime of the
lower mass stars, which outlive the pulsar.

These numbers provide upper limits to the fraction of runaways which
have a visible pulsar as companion as we did not take the limited
beaming of the radio signal into account.  The beaming fraction may be
about 1/3, decreasing $f_{\rm run}$ to $\aplt 0.3$\,\% for O type
runaways and to $\sim 0.6$\% for spectral type B stars.  Realistically
speaking we expect that about one in 350 runaway O stars and one in
200 B stars may be observable through the accompanying radio pulsar. A
considerably larger fraction may be observed as an X-ray
source. Runaway binaries in which the pulsar is visible have generally
rather large orbital periods to allow the pulsar to shine through the
dense wind of the O star. These binaries therefore tend to have
relatively small runaway velocity (see Fig.\,\ref{fig:Pv_bin}).

The high velocity stars of spectral type A (and later) found by Lance
(1988a) may in part be explained from binary supernova ejections.  Not
all their characteristics, however, are satisfactory explained from
this model.  These stars do not show abnormal high rotational
velocities and tend to appear considerably younger than the A stars in
the Galactic disc.  The latter characteristic indicates that these
stars are blue stragglers.  This is somewhat unexpected as the later
spectral type stars tend to spend less time as blue stragglers (see
Fig.\,\ref{fig:MfBss}).  Figure\,\ref{fig:Mrunfrac} shows that the
specific frequency of runaways drops for lower masses and we expect
that the number of high velocity A stars ejected by supernova
explosions is small. A mass ratio distribution which peaks to zero
increases the fraction of runaways among later spectral types
considerably.  Figure\,\ref{fig:Mv_all} on the other hand shows that
lower mass stars tend to receive higher runaway velocities, which is
consistent with the observations (Lance 1988b). It is, however,
arguable that the minority of late type stars obtain their high
velocities or large distance from the Galactic disc by supernovae.

\section{Conclusions}

We perform detailed binary population synthesis calculations to study
the origin of runaway O and B stars.  

The majority of the OB runaways can be explained with the evolution of
binaries if: 1) the mass ratio distribution for binaries with early
type primaries strongly peaks towards unity, 2) stars are
rejuvenated to zero age by mass accretion from a companion star in a
close binary, and 3) the binary fraction among early type stars is
close to 100\%. Taking these requirements into consideration we
conclude that about 30\% of the runaway O stars but the majority of
runaway B stars may originate from evolving binaries.

Runaway O and B stars which are ejected in the supernova explosion of
their binary companions have the following characteristics:
\begin{itemize}
\item[1.] Half the type O runaways and at least 10\% of the B stars
	  are observable as blue stragglers relative to the 
	  association in which they were born.
\item[2.] High velocity runaways have higher equatorial rotation and
	  lower surface helium abundances. The runaways in shorter
	  period binaries have higher space velocities.
\item[3.] The binary fraction among runaway B stars is about 20\%
	  and increases to about 40\% for early type O stars.
\item[4.] For late type B stars ($m\aplt8$\,\msun) the accompanying
	  pulsar is most likely dead. 
\item[5.] The radio signal of most pulsars in runaway binaries are
	  unobservable due to absorption in the stellar wind.  
	  For early B stars ($m\apgt7$\,\msun) only $\aplt2$\,\% of the
	  pulsars are visible, this fraction drops to $\aplt1$\,\% for
	  O star runaways ($m\apgt15$\,\msun). 
\end{itemize}

\vskip 1in\noindent{\em Acknowledgements:}~~ 
This work was supported by NASA through Hubble Fellowship grant
HF-01112.01-98A awarded by the Space Telescope Science Institute,
which is operated by the Association of Universities for Research in
Astronomy, Inc., for NASA under contract NAS\, 5-26555.


\figcaption[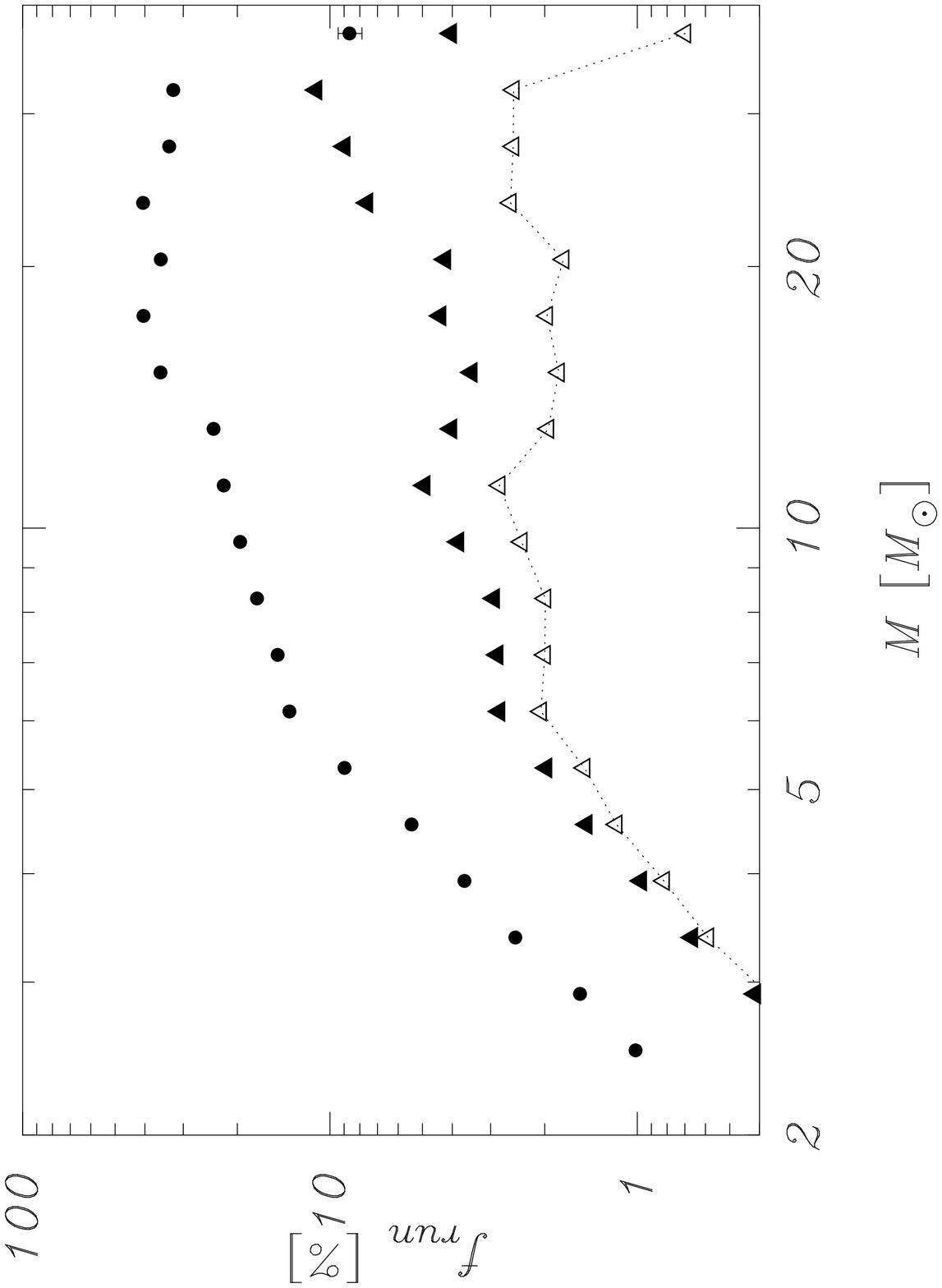] {Stars and binaries which are ejected
upon a supernova as fraction of the primaries initialized with the
same mass.  The filled symbols give number fractions, open symbols
represent observable fractions (see Eq.\,\ref{Eq:frun}). The latter
are corrected for the time that a star spends as a runaway.  The
fraction for stars which are ejected upon a supernova are identified
with $\bullet$, triangles gives the fraction of runaways among these
(stars with $v>25$\,\kms).  A Poissonian $1\sigma$ error bar is
presented at the far right bullet.
\label{fig:Mrunfrac}
}

\figcaption[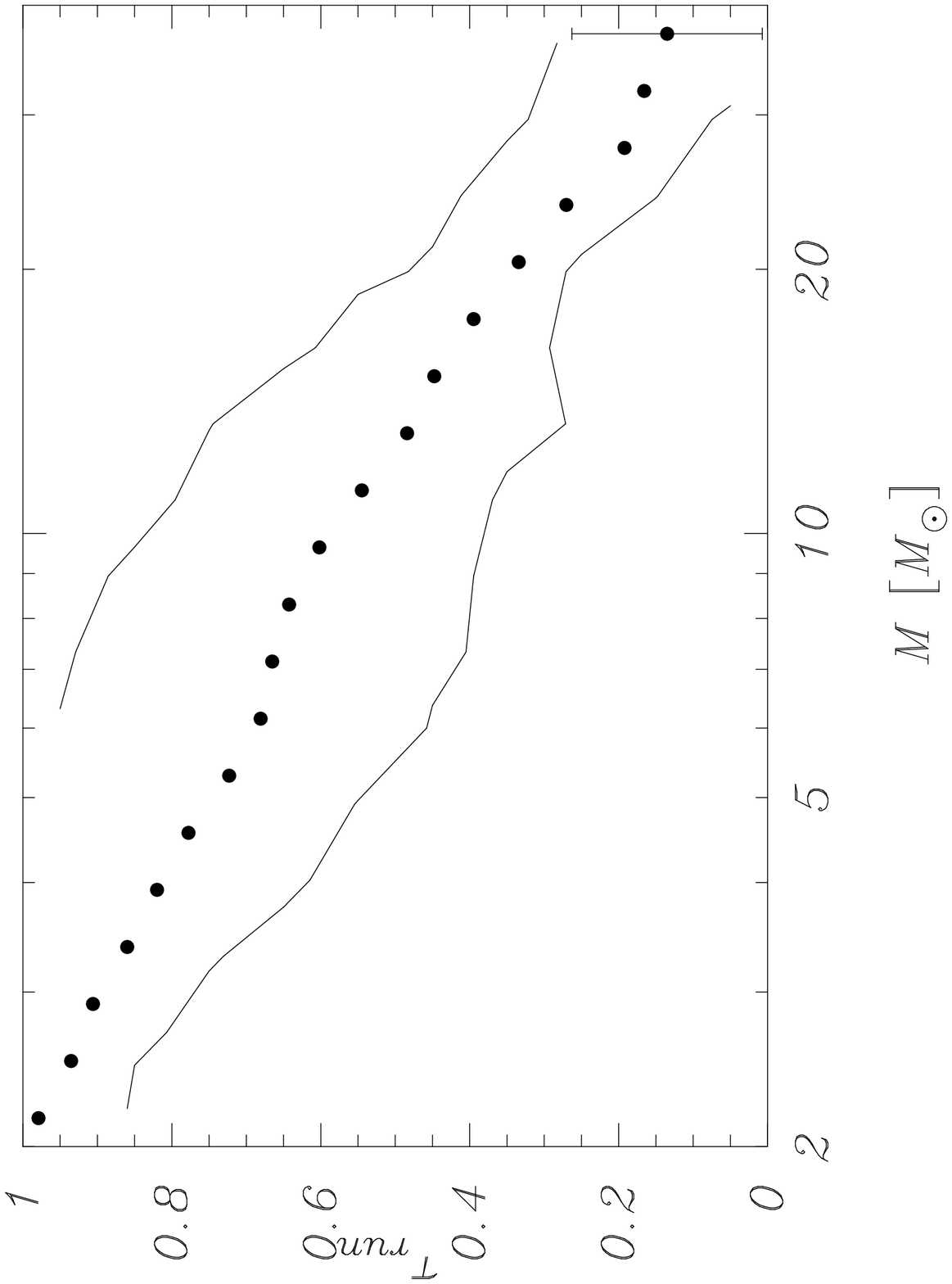] { Fraction of the lifetime
that a star spends as a runaway (see Eq.\,\ref{Eq:taurun}).  The
distribution is broad, bullets give the mean for the specific mass
bins, solid lines give the 90\% confidence intervals and the error bar
to the most right point gives the dispersion of the distribution.
\label{fig:Mtrunaway}}

\figcaption[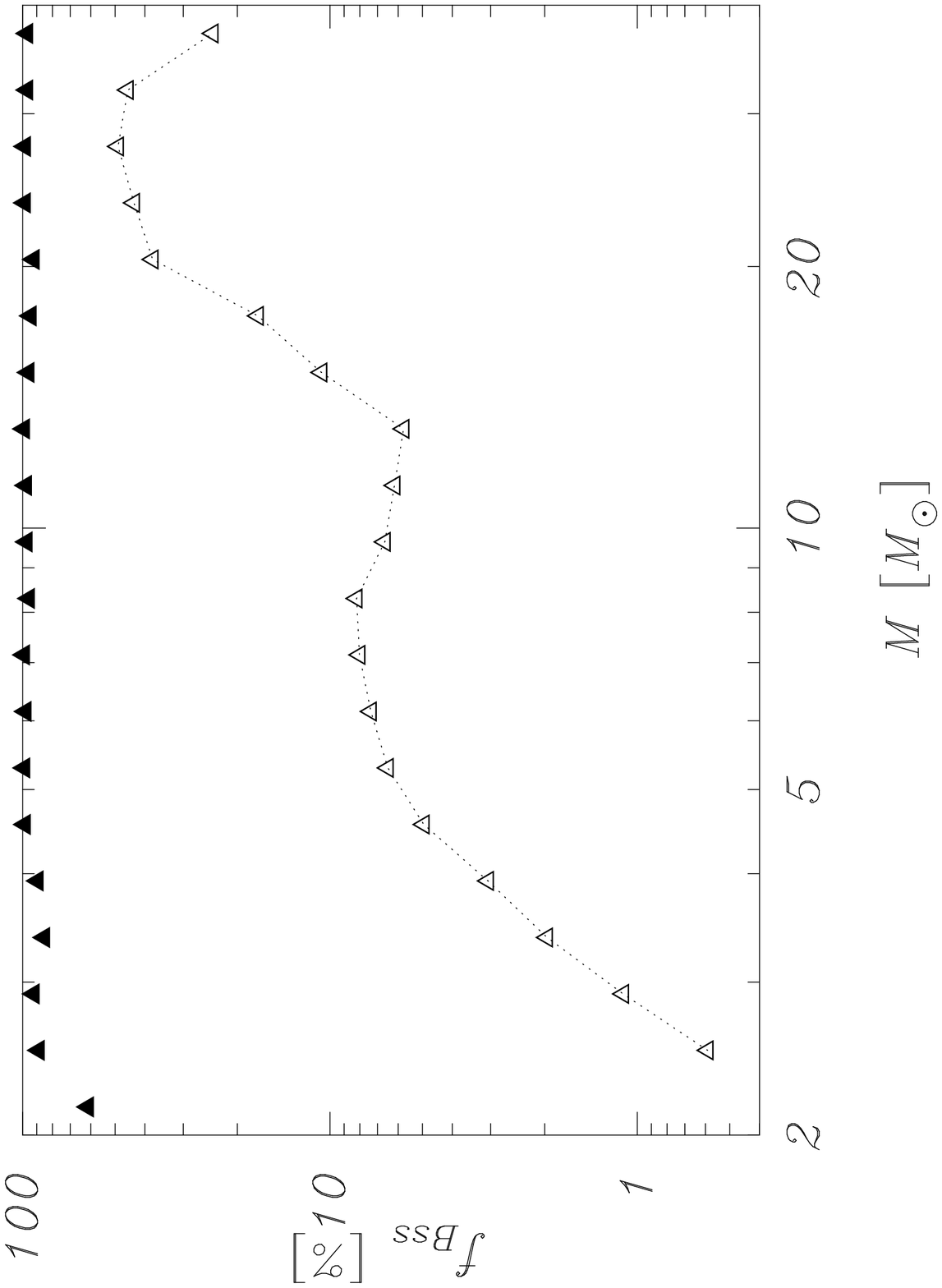] {Number fraction of blue stragglers
among runaways (filled triangle).  These fractions do not incorporate
the time a runaway spends as a blue straggler.  The open triangles
correct for this by weighting the time a star spend as a runaway by
the time it is observable as a blue straggler.
\label{fig:MfBss}}

\figcaption[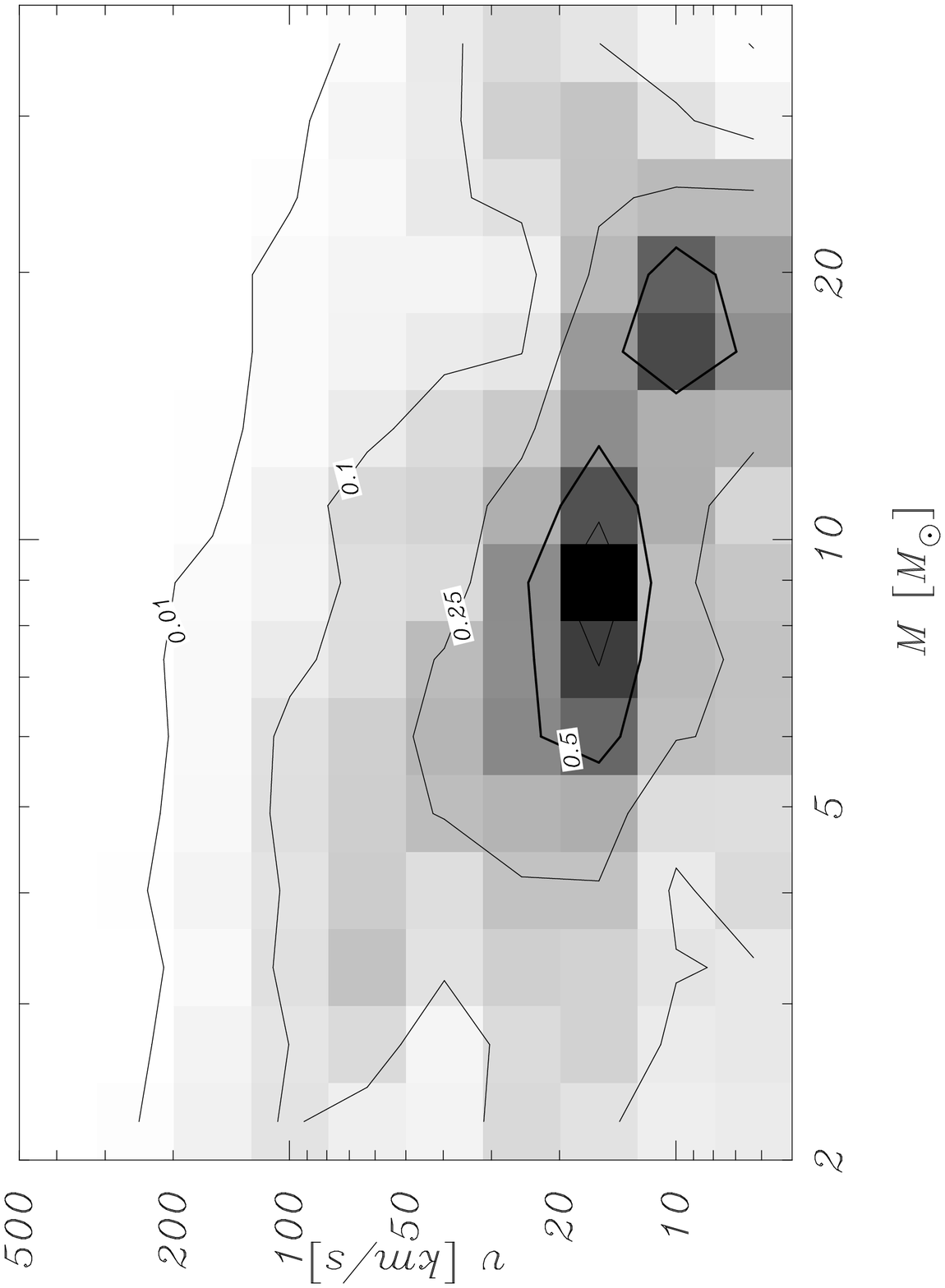] {Number of stars and binaries that are
ejected upon a supernova (gray shades and contours) as a function of
velocity and the mass of the runaway (the visible component in the
case of a binary). Gray shades are linear in number density.
\label{fig:Mv_all}
}


\figcaption[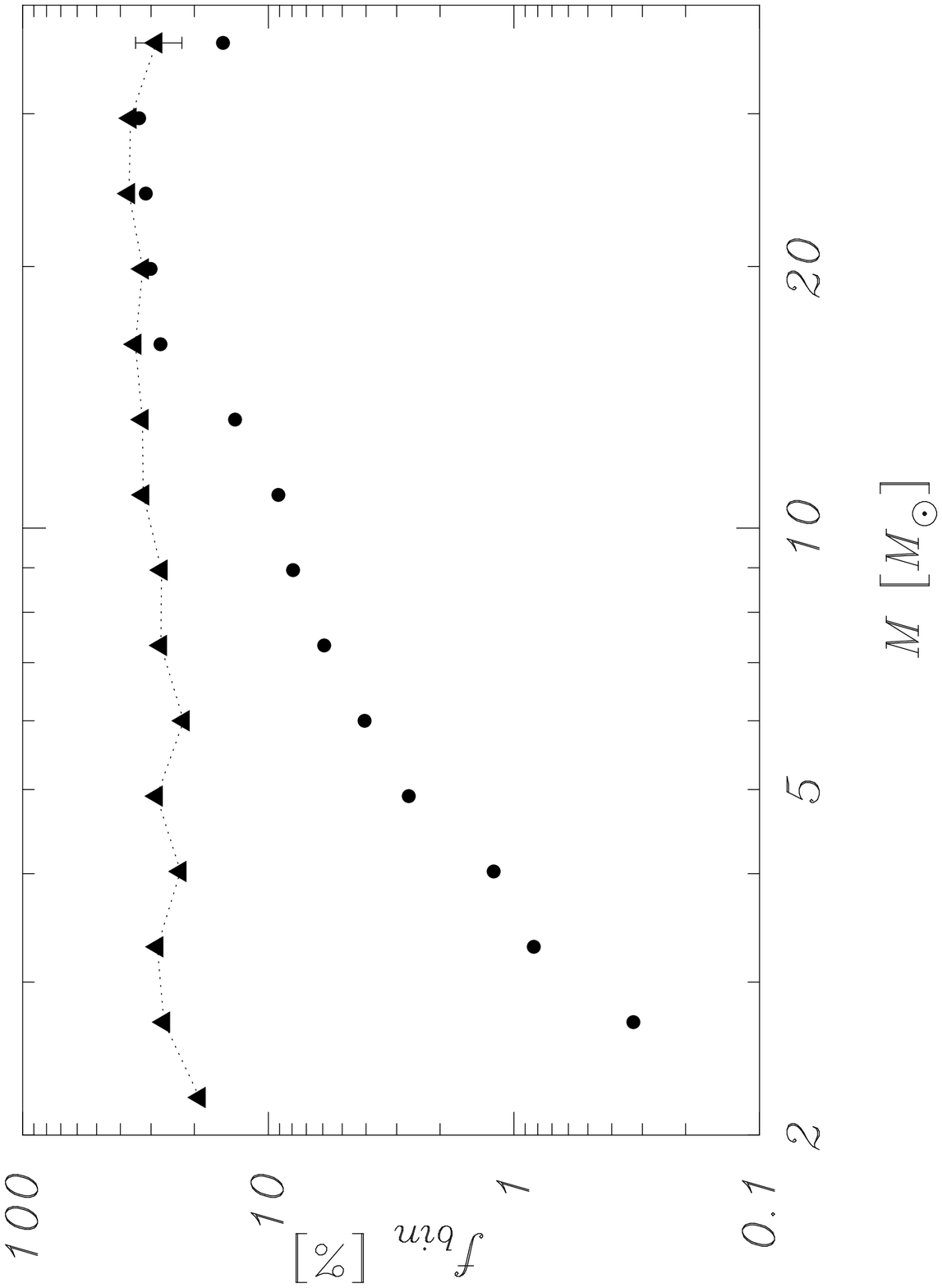] {Binary fraction as a
function of the mass of the observable component. The bullets indicate
the fraction of binaries which survive the supernova.  The triangles
give the binary fraction among runaways ($v>25$\,\kms).  Error bars
are $1\sigma$ Poissonian.
\label{fig:Mbinfrac}
}

\figcaption[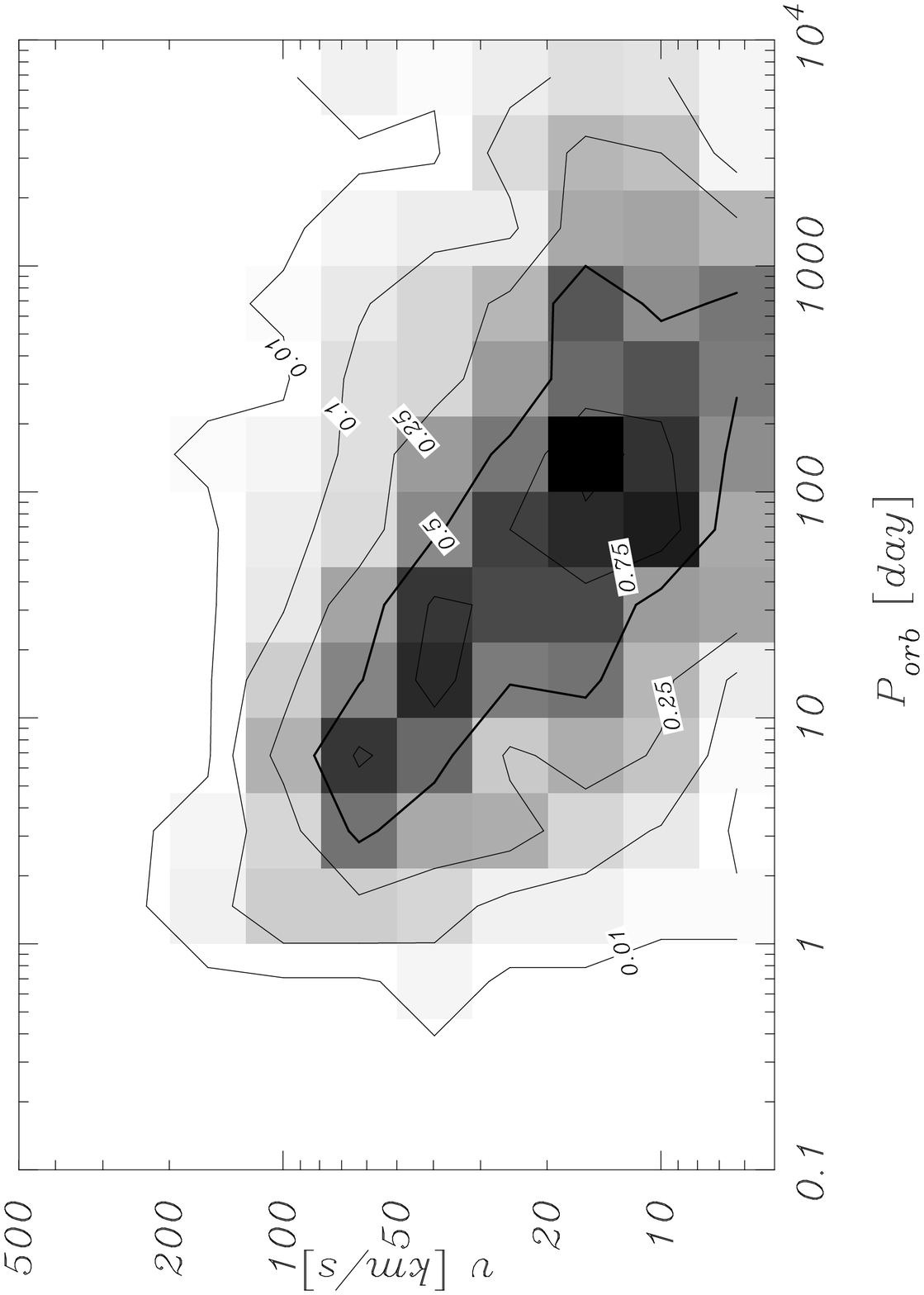] {Probability distribution of runaway
velocity as a function of orbital period for binaries with a visible
star of $m>5$\,\msun.
\label{fig:Pv_bin}}

\figcaption[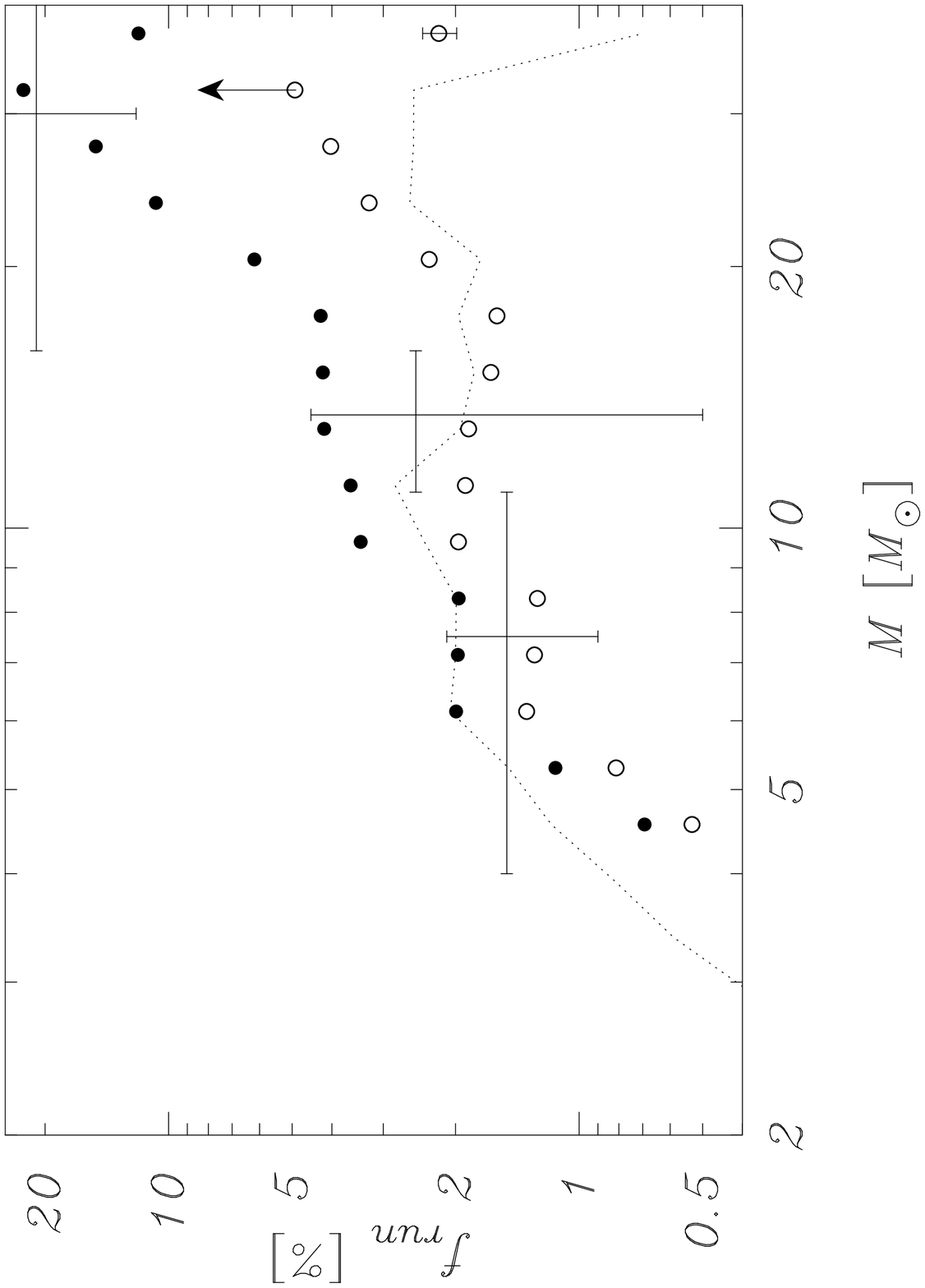] {Specific frequency of runaway stars
weighed with lifetime.  The dotted line give the results of the
standard model (open triangles in Fig.\,\ref{fig:Mrunfrac}).  Circles
give the results for a mass ratio distribution which is peaked to
unity. The arrow indicates the increase in the fraction of runaways
when the maximum semi major axis is reduced by two orders of magnitude
to $10^4$\,\rsun.  The $\bullet$ show the specific frequency
of runaways assuming that an accreting star is rejuvenated to zero
age.  The error bar to the last $\circ$ shows the $1 \sigma$\,
Poissonian error. The other error bars indicate the observed fraction
or runaways from Blaauw (1961; 1993).
\label{fig:Mrunaway}}

\figcaption[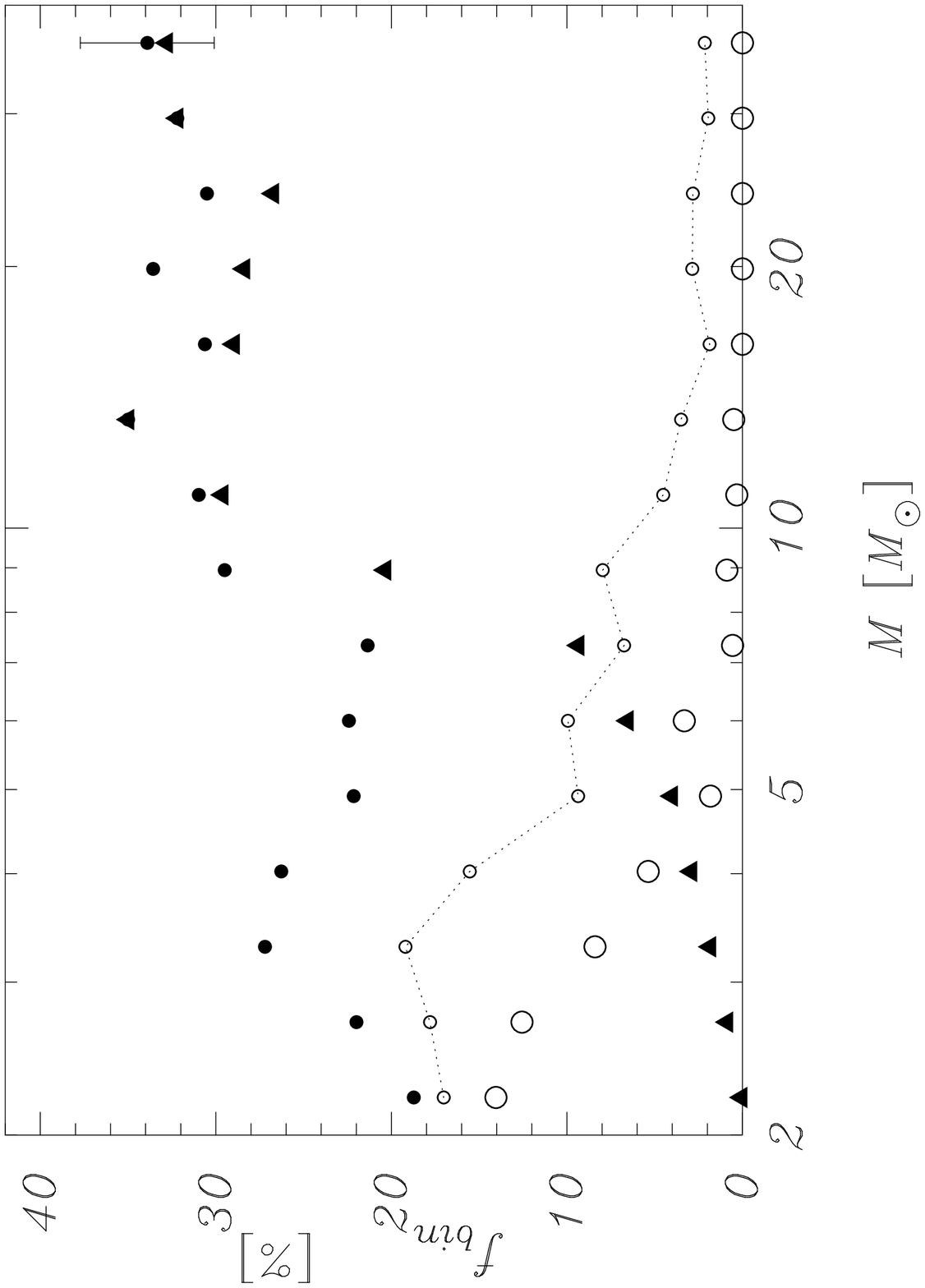] {Binary fraction among runaway
stars as a function of mass ($\bullet$, see also
fig.\,\ref{fig:Mbinfrac}).  The filled triangles give the fraction of
binaries in which the neutron star is active as a radio pulsar
(weighted on the lifetime of the runaway star). The small open circles
(connected with the dotted line) gives the specific frequency of runaway
binaries for which the pulsar is visible through the stellar wind at
apocenter, the large circles show the fraction for which the pulsar is
visible all the time.
\label{fig:MvisPbinfrac}}

\clearpage
\begin{deluxetable}{rrrrr}
\tablecaption{
Initial conditions for the binaries.
The first and second columns give the parameter and the
functionality, followed by the adopted lower and upper.}
\tablewidth{0pt}
\tablehead{
\colhead{parameter}& \colhead{function}     & \multicolumn{2}{c}{limits} \\
                   &                        & \colhead{lower} & \colhead{upper}\\
}
\startdata
mass function      & $P(M) = $\,Scalo (1986)& 1\,\msun    & 100\,\msun \\
secondary mass     & $P(m) = {\rm constant}$& 0.1\,\msun  & $M/\msun$ \\
orbital separation & $P(a) = 1/a$           & 1\,\rsun    & $10^6$\,\rsun \\
eccentricity       & $P(e) = 2e$            & 0           & 1    \\ 
\enddata
\label{Tab:Binit}
\end{deluxetable}

\clearpage
\plotone{fig/fig_Mrunfrac.ps}
\clearpage
\plotone{fig/fig_mtrunaway.ps}
\clearpage
\plotone{fig/fig_MftBss.ps}
\clearpage
\plotone{fig/fig_Mv_all.ps}
\clearpage
\plotone{fig/fig_Mbinfrac.ps}
\clearpage
\plotone{fig/fig_Pv_bin.ps}
\clearpage
\plotone{fig/fig_frunaway.ps}
\clearpage
\plotone{fig/fig_MvisPbinfrac.ps}

\end{document}